\documentclass[twocolumn,floatfix,aps]{revtex4-2}
\usepackage{graphicx}
\usepackage{amsmath}
\usepackage{amssymb}
\usepackage{bm}
\usepackage{hyperref}

\begin{document}

\title{Pure dephasing increases partition noise in the quantum Hall effect}
\author{C. W. J. Beenakker}
\affiliation{Instituut-Lorentz, Universiteit Leiden, P.O. Box 9506, 2300 RA Leiden, The Netherlands}

\date{September 2025}

\begin{abstract}
Quantum Hall edge channels partition electric charge over $N$ chiral (uni-directional) modes. Intermode scattering leads to partition noise, observed in graphene \textit{p-n} junctions. While inelastic scattering suppresses this noise by averaging out fluctuations, we show that pure (quasi-elastic) dephasing may enhance the partition noise. The noise power increases by up to 50\% for two modes, with a general enhancement factor of $1+1/N$ in the strong-dephasing limit. This counterintuitive effect is explained in the framework of monitored quantum transport, arising from the self-averaging of quantum trajectories.
\end{abstract}
\maketitle

\begin{figure}[tb]
\centerline{\includegraphics[width=0.7\linewidth]{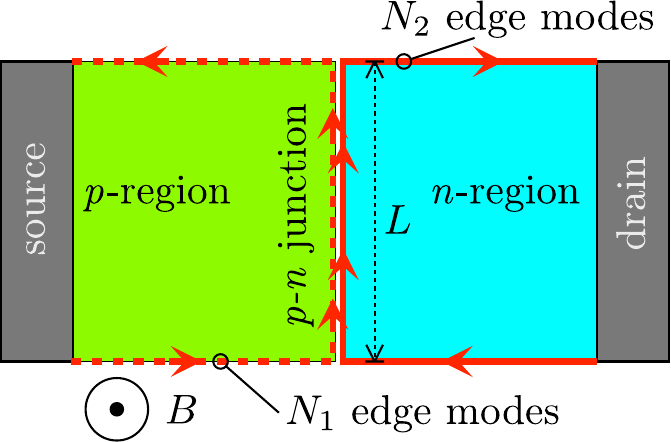}}
\caption{Electronic beam splitter in graphene \cite{Wil07,Aba07}. A potential step separates a region where the Fermi level lies in the lower half of the Dirac cone (\textit{p}-region) from a region where it lies in the upper half (\textit{n}-region). Quantum Hall edge modes circulate around each region, in opposite directions (dashed and solid arrows). The edge modes are merged and mixed when they enter one end of the \textit{p-n} junction, until they are again split at the other end. A voltage bias $V$ injects electrons from the source into $N_1$ modes in the \textit{p}-region, a fraction ${\cal T}$ of which is transferred to the drain via $N_2$ modes in the \textit{n}-region. The discreteness of the electron charge produces partition noise, modified by quantum interference effects.
}
\label{fig_layout}
\end{figure}

\textbf{\em Introduction ---}
Partition noise is generated when a flow of discrete particles must split into different paths, producing time dependent current fluctuations (a form of shot noise \cite{Blanter00}). In quantum electronics, this effect of the electron as a particle is modified by its wave nature via quantum interference of the different paths \cite{Boc12}. This wave-particle interplay has been studied extensively in graphene \cite{Wil07,Aba07,Two07,Li08,Vel09,Car10,Car11,Che11,Mat15,Kum15,Fra16,Han17,Ma18,Kum18,Zeb18}, where a \textit{p-n} junction in a magnetic field functions as an electronic beam splitter (see Fig.\ \ref{fig_layout}). 

The relevance of quantum interference can be controlled by the number $N=N_1+N_2$ of electronic modes that are partitioned. In graphene these are quantum Hall edge states propagating unidirectionally, chirally, along the junction. If the chiral modes are fully mixed by phase coherent random scattering, the expectation value $\mathbb{E}[P_{\rm coh}]$ of the zero-temperature noise power  is \cite{Sav06,Bra06}
\begin{equation}
\mathbb{E}[ P_{\rm coh}]=P_0\frac{N_1^2 N_2^2}{N(N^2-1)}.\label{Pcoherent}
\end{equation} 
Here $P_0=eVG_0$ with $V$ the bias voltage and $G_0$ the conductance quantum ($G_0=e^2/h$ for non-degenerate modes). In the case $N_1=N_2$ of equal partitioning, the average of the noise power per mode $N_1^{-1}P_{\rm coh}/P_0$ decreases from $1/6$ at $N_1=1$ to $1/8$ in the semiclassical large-$N$ limit \cite{Jal94,Bla00,Obe01}.

One might surmise from this $N$-dependence that loss of phase coherence \textit{reduces} the partition noise. Inelastic scattering certainly has that effect, it may fully suppress the current fluctuations \cite{Ma18}. Contrary to this intuition, we find that pure dephasing, without significant inelastic scattering, may \textit{increase} the partition noise.

\textbf{\em Monitored quantum transport ---}
Fully phase-coherent dynamics of $N$ chiral modes is described by an $N\times N$ unitary matrix, which for strong intermode scattering may be assumed to be distributed uniformly in the unitary group \cite{Aba07}. The statistics of the transferred charge is then governed by the circular unitary ensemble (CUE) of random-matrix theory \cite{Bee97}. We will explore how the charge transfer statistics is modified if the coherent dynamics is combined with weak measurements. This is an application of the method of ``monitored quantum transport'', which I recently developed with Jin-Fu Chen \cite{Bee25}.

Our interest is in ``pure'' dephasing, a quasi-elastic process with negligible energy transfer. Coherence is destroyed when electrons becomes entangled with the environment, which effectively measures (``monitors'') the electron wave function. The dephasing is quasi-elastic if the characteristic frequency $\omega_c$ of fluctuations in the environment (for example, gate voltage fluctuations) is large compared to the applied voltage $V$, but small compared to the inverse dwell time $t_{\rm dwell}={L}/v_{\rm F}$ of an electron in the junction (of length ${L}$) \cite{Mar04,notedephasing}.

In the framework of monitored quantum transport we implement pure dephasing by projective measurements in the energy eigenbasis. We consider a single-mode projection $|n\rangle\langle n|$, for some arbitrary mode $n$. The measurements alternate with unitary mixing of the modes. Since a change in the measured mode index can be absorbed by a basis change of the unitaries, we may without loss of generality assume that it is always the same mode $n$ that is measured.

To have a variable dephasing strength, we allow for a ``weak'' measurement, which interpolates with weight $\varepsilon\in(0,1)$ between the identity $\hat{I}$ and a projection onto a filled or empty state,
\begin{equation}
\begin{split}
&\hat{P}_{+}=\delta\hat{I} +\varepsilon a_n^\dagger a_n^{\vphantom{\dagger}}=\delta e^{\gamma a_n^\dagger a_n^{\vphantom{\dagger}}},\\
&\hat{P}_{-}=\delta\hat{I}+\varepsilon a_n^{\vphantom{\dagger}}a_n^\dagger=\delta e^\gamma e^{-\gamma a_n^\dagger a_n^{\vphantom{\dagger}}},\\
&\delta=\tfrac{1}{2}(\sqrt{2-\varepsilon^2}-\varepsilon),\;\;\gamma=\ln(1+\varepsilon/\delta).
\end{split}
\end{equation}
The operators $a_n^\dagger$, $a_n$ are fermionic creation and annihilation operators. The coefficient $\delta$ is chosen such that 
\begin{equation}
\hat{P}_{+}^2+\hat{P}_{-}^2=\hat{I}.\label{Pidentity}
\end{equation}

The monitored transport along the junction is decomposed into $\cal L$ alternations of unitary mixing and weak measurements. Unitary mixing in segment $\ell$ is described by the matrix $U_{\ell}\in{\rm U}(N)$ of scattering amplitudes. In second quantization, the corresponding scattering operator has the Gaussian form
\begin{equation}
\hat{U}_\ell=e^{ia^\dagger M_\ell a},\;\;e^{iM_\ell}=U_\ell.
\end{equation}
We have collected the fermion operators in a vector, $a=(a_1,a_2,\ldots a_N)$, contracted with the Hermitian $N\times N$ matrix $M_\ell$.

\begin{figure}[tb]
\centerline{\includegraphics[width=0.8\linewidth]{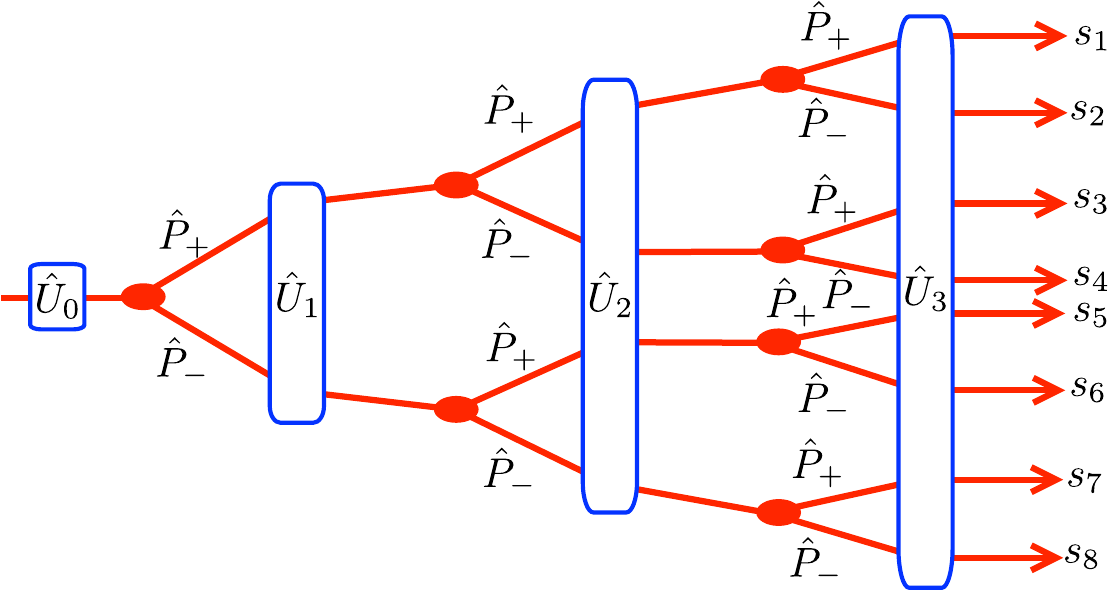}}
\caption{Monitored quantum transport model of dephasing \cite{Bee25}: Phase coherent mixing of the modes (unitary operators $\hat{U}_i$) alternates with weak measurements of the occupation number of a mode (operators $\hat{P}_+,\hat{P}_-$, depending on whether the mode is found filled or empty). The string $s_i$ of ${\cal L}$ measurement outcomes selects one of the $2^{\cal L}$ quantum trajectories (here shown for ${\cal L}=3$). The corresponding Kraus operators \eqref{Krausdef} are correlated because they contain the same set of unitary operators.
}
\label{fig_trajectories}
\end{figure}

The measurement outcome of mode $n$ in segment $\ell$ can be either ``filled'' ($s_\ell=+$) or ``empty'' ($s_\ell=-$). The string of measurement outcomes $\bm{s}=\{s_{\cal L},s_{{\cal L}-1},\ldots s_1\}$ labels the Kraus operator \cite{NielsenChuang}
\begin{equation}
\hat{K}_{{\bm{s}}}=\hat{U}_{{\cal L}}\hat{P}_{s_{\cal L}}\hat{U}_{{\cal L}-1}\hat{P}_{s_{{\cal L}-1}}\hat{U}_{{\cal L}-2}\hat{P}_{s_{{\cal L}-2}}\cdots \hat{U}_1\hat{P}_{s_1}\hat{U}_0,\label{Krausdef}
\end{equation}
which governs the evolution of the density matrix,
\begin{equation}
{\rho}_{\rm out}=\textstyle{\sum_{\bm{s}} }\hat{K}_{\bm{s}} \rho_{\rm in}\hat{K}^\dagger_{\bm{s}} .\label{mapping}
\end{equation}
An individual term in this operator sum is referred to as a ``quantum trajectory'', see Fig.\ \ref{fig_trajectories}.

The mapping \eqref{mapping} is trace preserving in view of the sum rule
\begin{equation}
\textstyle{\sum_{\bm{s}}}\hat{K}_{\bm{s}}^\dagger \hat{K}_{\bm{s}}^{\vphantom{\dagger}}=\hat{I},\label{sumrule}
\end{equation}
guaranteed by Eq.\ \eqref{Pidentity}. A Gaussian density matrix is mapped onto a convex sum of Gaussian operators. Such a quantum channel is called a mixed Gaussian channel \cite{Mel13}. The channel is ``unital'', meaning that the identity is mapped onto itself \cite{note1}, which is characteristic for dephasing without dissipation. 

\textbf{\em Charge transfer statistics ---}
Charge is injected by a voltage bias $V$ into $N_{1}$ incident modes and detected during a time $t_{\rm counting}$ in $N_{2}$ outgoing modes. We seek the moment generating function $F(\xi)=\operatorname{Tr}\rho_{\rm out}e^{\xi{Q}}$ of the transferred charge $Q$ (in units of the electron charge). At low temperatures, $k_{\rm B}T\ll eV$, and in the long-time limit, ${\cal N}_V\equiv eVt_{\rm counting}/h\gg 1$, this is given by a quantum channel generalization \cite{Bee25} of the Levitov-Lesovik formula \cite{Lev93,Lev96},
\begin{subequations}
\label{Fxideterminant}
\begin{align}
& F(\xi)=\bigl[\textstyle{\sum_{\bm s}}F_{\bm s}(\xi)\bigr]^{{\cal N}_V},\\
& F_{\bm s}(\xi)=c_{\bm s}^2\operatorname{Det}\bigl(1+P_{\rm in}[{K}_{\bm s}^\dagger e^{\xi P_{\rm out}}{K}_{\bm s}-1]\bigr).\label{Fxideterminantb}
\end{align}
\end{subequations}
Here $P_{\rm in},P_{\rm out}$ project onto the $N_{1},N_{2}$ incoming and outgoing modes, and we have defined the Kraus matrix
\begin{subequations}
\label{Krausmatrixdef}
\begin{align}
&{K}_{\bm s}=U_{{\cal L}}{P}_{s_{\cal L}}{U}_{{\cal L}-1}{P}_{s_{{\cal L}-1}}{U}_{{\cal L}-2}{P}_{s_{{\cal L}-2}}\cdots {U}_1{P}_{s_1}{U}_0,\\
&P_{\pm}=e^{\pm\gamma |n\rangle\langle n|},\;\;c_{\bm s}=\delta^L e^{\sum_\ell (1-s_\ell)\gamma/2}.
\end{align}
\end{subequations}

The unitary matrices $U_{\ell}$ are the single-particle scattering matrices at the Fermi level, assumed to be nearly energy independent on the scale of $eV$. This applies if $eVt_{\rm dwell}/\hbar\ll 1$, so together with the quasi-elastic requirement we work in the regime
\begin{equation}
eV\ll \hbar\omega_c\ll \hbar/t_{\rm dwell}.
\end{equation}

The Krauss matrix ${K}_{\bm s}$ is not unitary \cite{note2}. If we rescale
\begin{equation}
{K}_{\bm{s}}=e^{(\gamma/N)\sum_{\ell=1}^{\cal L} s_\ell}{\cal S}_{\bm s},\label{KSrelation}
\end{equation}
the matrix ${\cal S}_{\bm s}$, while still non-unitary, is unimodular: $|\operatorname{Det} S_{\bm s}|=1$. It is convenient to partition it into transmission and reflection blocks,
\begin{equation}
{\cal S}_{\bm s}=\begin{pmatrix}
r_{\bm{s}}&t'_{\bm{s}}\\
t_{\bm{s}}&r'_{\bm{s}}
\end{pmatrix},\;\;
P_{\rm in}=\begin{pmatrix}
1&0\\
0&0
\end{pmatrix},\;\;P_{\rm out}=\begin{pmatrix}
0&0\\
0&1
\end{pmatrix}.\label{Spartioning}
\end{equation}
The dimensions of $r_{\bm s}$ and $t_{\bm s}$ are $N_1\times N_1$ and $N_2\times N_1$, respectively.

With this decomposition we may write the moment generating function \eqref{Fxideterminant} as a sum over determinants of $N_1\times N_1$ Hermitian matrices \cite{note5},
\begin{align}
&F(\xi)=\bigl[\textstyle{\sum_{\bm s}}C_{\bm s}\operatorname{Det}\bigl(r_{\bm s}^\dagger r_{\bm s}^{\vphantom{\dagger}}+e^\xi t_{\bm s}^\dagger t_{\bm s}^{\vphantom{\dagger}}\bigr)\bigr]^{{\cal N}_V},\;\;C_{\bm s}=C_{\cal L}\Delta_{\bm s},\nonumber\\
&C_{\cal L}=(2\cosh\gamma)^{-\cal L},\;\;\Delta_{\bm s}=e^{(N_1-N_2)(\gamma/N)\sum_\ell s_\ell}.
\label{Fxidetrt}
\end{align}
Without the measurements, for $\gamma=0$, the matrix ${\cal S}_{\bm s}\equiv {\cal S}_0$ is unitary, there is a single $\bm{s}$, and one recovers the Levitov-Lesovik formula  \cite{Lev93,Lev96}
\begin{equation}
F_0(\xi)=\operatorname{Det}[1+(e^{\xi}-1)t_0^\dagger t_0^{\vphantom{\dagger}}]^{{\cal N}_V}.
\end{equation}

Normalization, $F(0)=1$, is ensured by the sum rule
\begin{equation}
\textstyle{\sum_{\bm s}}p_{\bm s}=1,\;\;p_{\bm s}=C_{\bm s}\operatorname{Det}\bigl(r_{\bm s}^\dagger r_{\bm s}^{\vphantom{\dagger}}+t_{\bm s}^\dagger t_{\bm s}^{\vphantom{\dagger}}\bigr).\label{sumrule2}
\end{equation}
The number $p_{\bm s}\in(0,1)$ is the weight of the quantum trajectory, meaning it is the Born-rule probability of one of the $2^{\cal L}$ measurement outcomes labeled by the string $\bm s$. The moment generating function \eqref{Fxidetrt} can be rewritten as  a Born-weighted sum over a quantum trajectory dependent transfer probability ${\cal T}_{\bm s}$,
\begin{equation}
\begin{split}
&F(\xi)=\bigl[\textstyle{\sum_{\bm s}}p_{\bm s}\operatorname{Det}\bigl(1+(e^\xi-1) {\cal T}_{\bm s}\bigr)\bigr]^{{\cal N}_V},\\
&{\cal T}_{\bm s}=(r_{\bm s}^\dagger r_{\bm s}^{\vphantom{\dagger}}+t_{\bm s}^\dagger t_{\bm s}^{\vphantom{\dagger}})^{-1}t_{\bm s}^\dagger t_{\bm s}^{\vphantom{\dagger}}.
\end{split}
\label{FxipsTs}
\end{equation}

\textbf{\em Single-mode case ---}
To simplify the calculation we first specialize to the single-mode case $N_{1}=N_{2}=1$. (The multi-mode case of arbitrary $N_1,N_2$ is adressed later on.) In the graphene \textit{p-n} junction of Fig.\ \ref{fig_layout} this applies to a single valley-degenerate Landau level without spin-orbit coupling, the spin degree of freedom is absorbed in $G_0=2e^2/h$. We label the incoming mode in the \textit{p}-region by 1 and the outgoing mode in the \textit{n}-region by 2. 

The length ${L}$ of the junction is set at a large multiple $\cal L$ of the length $L_{\rm mixing}$ on which the two modes are fully mixed by elastic intervalley scattering. The dephasing length $L_\phi$ is greater than the mixing length,
\begin{equation}
L_\phi=\varepsilon^{-2} L_{\rm mixing}\Rightarrow \varepsilon^2{\cal L}=L/L_\phi.
\end{equation}
In the regime $L,L_\phi> L_{\rm mixing}$ we may draw the matrices $U_{\ell}$ uniformly (with the Haar measure \cite{Bee97}) from ${\rm U}(2)$.

The transmission matrix $t_{\bm s}$ (from mode 1 to mode 2) and the reflection matrix $r_{\bm s}$ (from mode 1 to mode 1) now each consist of a single complex number, and $C_{\bm s}=C_{\cal L}$, so the moment generating function \eqref{Fxidetrt} simplifies to
\begin{align}
F(\xi)={}&\bigl[\textstyle{\sum_{\bm s}}C_{\cal L}\bigl(|r_{\bm s}|^2+e^\xi |t_{\bm s}|^2\bigr)\bigr]^{{\cal N}_V}\nonumber\\
={}&[1+(e^\xi-1){\cal T}]^{{\cal N}_V},\;\;{\cal T}=C_{\cal L}\textstyle{\sum_{\bm s}} |t_{\bm s}|^2,\label{FxiNis2}
\end{align}
where we have used the sum rule \eqref{sumrule2}. Eq.\ \eqref{FxiNis2} describes binomial statistics, a general property of single-mode detection \cite{Bee25}. The transfer probability ${\cal T}$ determines the conductance $G$ and the noise power $P_{\rm noise}$ \cite{Blanter00},
\begin{equation}
G=G_0{\cal T},\;\;P_{\rm noise}=P_0{\cal T}(1-{\cal T}).
\end{equation}

\textbf{\em Statistics of a single quantum trajectory ---}
For insight, it is helpful to first consider a single quantum trajectory. If one could post-select the data based on a particular string $\bm{s}$ of measurement outcomes, one would observe a transfer probability
\begin{equation}
{\cal T}_{\bm s}=\frac{C_{{\cal L}}|t_{\bm s}|^2}{p_{\bm s}}=\frac{|t_{\bm s}|^2}{|r_{\bm s}|^2+|t_{\bm s}|^2}.
\end{equation}

To find the statistics of ${\cal T}_{\bm s}$, we use that ${\cal S}_{\bm s}$ can be factored as ${\cal S}_{\bm s}=UM$ with $U$ uniformly distributed in ${\rm U}(2)$, independently of the $2\times 2$ complex matrix $M$. The absolute value squared of any matrix element of $U$ is uniformly distributed in $(0,1)$ \cite{Bee97,note4}. Denote the first column of $M$ by the vector ${\bm m}=(m_1,m_2)^\top$, then $({r}_{\bm s},{t}_{\bm s})^\top=U\cdot\bm{m}$, hence
\begin{equation}
{\cal T}_{\bm s}=\frac{|U_{21}m_1+U_{22}m_2|^2}{|m_1|^2+|m_2|^2}. 
\end{equation}
Because the Haar measure is invariant under multiplication by an arbitrary fixed unitary, we may rotate $\bm{m}$ so that $m_2=0$, without changing the probability distribution of ${\cal T}_{\bm s}$ in the circular unitary ensemble of random $U$. Now the vector $\bm{m}$ drops out of ${\cal T}_{\bm s}=|U_{21}|^2$, which therefore retains the uniform distribution in $(0,1)$.

\begin{figure}[tb]
\centerline{\includegraphics[width=0.8\linewidth]{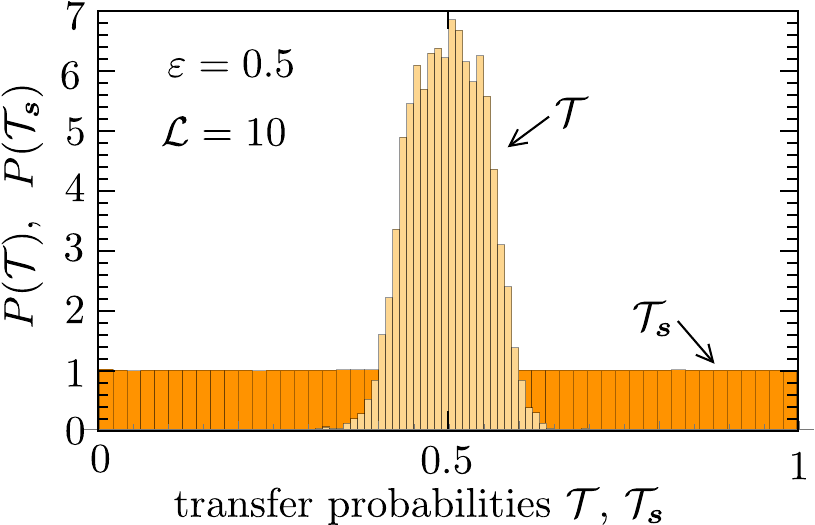}}
\caption{Distribution of the transfer probability in the ensemble of random unitary matrices [Haar-measure distributed in ${\rm U}(N)$ with $N=2$]. The transfer probability ${\cal T}_{\bm s}$ of a single quantum trajectory is uniformly distributed in $(0,1)$, while the full transfer probability ${\cal T}$, summed over all quantum trajectories, is narrowly peaked around $1/2$. The histograms \cite{note_numerics} are computed numerically from the Kraus matrix \eqref{Krausmatrixdef}, averaged over $5000$ sets of unitaries $U_0,U_1,\ldots U_{\cal L}$, for $\varepsilon=0.5$ and ${\cal L}=10$.
}
\label{fig_ptransfer}
\end{figure}

We have thus found that the statistics of the transfer probability ${\cal T}_{\bm s}$ of a single quantum trajectory is independent of the measurement outcome or measurement strength. The dependence of the observable transfer probability ${\cal T}=\sum_{\bm{s}}p_{\bm s}{\cal T}_{\bm s}$ on the measurements arises from the sum over the measurement outcomes with Born weight $p_{\bm s}$. This ``self-averaging'' converts the uniform distribution on $(0,1)$ of ${\cal T}_{\bm s}$ into a narrow distribution around 1/2 of ${\cal T}$. A numerical calculation shows the collapse of the distribution in Fig.\ \ref{fig_ptransfer}.

\textbf{\em Dephasing-induced noise increase ---}
In App.\ \ref{app_secondmoment} all moments of ${\cal T}$ are calculated, in the ensemble of random unitaries. For the average noise power we need the first two moments. The mean is $\mathbb{E}[{\cal T}]=1/2$, independent of the measurement strength. The variance is
\begin{equation}
\operatorname{Var}[{\cal T}]=\tfrac{1}{12}\left(1-\tfrac{4}{3}\varepsilon^2+\tfrac{2}{3}\varepsilon^4\right)^{\cal L}\rightarrow \tfrac{1}{12}e^{-4\varepsilon^2{\cal L}/3}\label{VarTresult}
\end{equation}
in the limit $\varepsilon\rightarrow 0$, ${\cal L}\rightarrow\infty$ at fixed $\varepsilon^{2}{\cal L}$. In this limit the noise power has average
\begin{equation}
\mathbb{E}[P_{\rm noise}]=P_0\bigl(\tfrac{1}{4}-\tfrac{1}{12}e^{-4\varepsilon^2{\cal L}/3}\bigr),
\end{equation}
increasing by up to 50\% (from $P_0/6$ to $P_0/4$) as dephasing becomes stronger and stronger.

\textbf{\em Multi-mode case ---}
Turning now to arbitrary $N_1,N_2$, we start from the general expression \eqref{FxipsTs} of the moment generating function, expand $F(\xi)=1+\langle Q\rangle\xi+\tfrac{1}{2}\langle Q^2\rangle\xi^2+{\cal O}(\xi^3)$,  to obtain the conductance $G=(G_0/{\cal N}_V)\langle Q\rangle$ and the noise power $P_{\rm noise}=(P_0/{\cal N}_V)[\langle Q^2\rangle-\langle Q\rangle^2]$. This results in 
\begin{align}
G/G_0={}&\textstyle{ \sum_{\bm s}}p_{\bm s}\operatorname{Tr}{\cal T}_{\bm s},\\
P_{\rm noise}/P_0={}&\textstyle{\sum_{\bm s}}p_{\bm s}\bigl[\operatorname{Tr}{\cal T}_{\bm s}(1-{\cal T}_{\bm s}) +(\operatorname{Tr}{\cal T}_{\bm s})^2\bigr]\nonumber\\
&-\bigl[\textstyle{ \sum_{\bm s}}p_{\bm s}\operatorname{Tr}{\cal T}_{\bm s}\bigr]^2.\label{Pnoiseformulasumps}
\end{align}

\begin{figure}[tb]
\centerline{\includegraphics[width=0.8\linewidth]{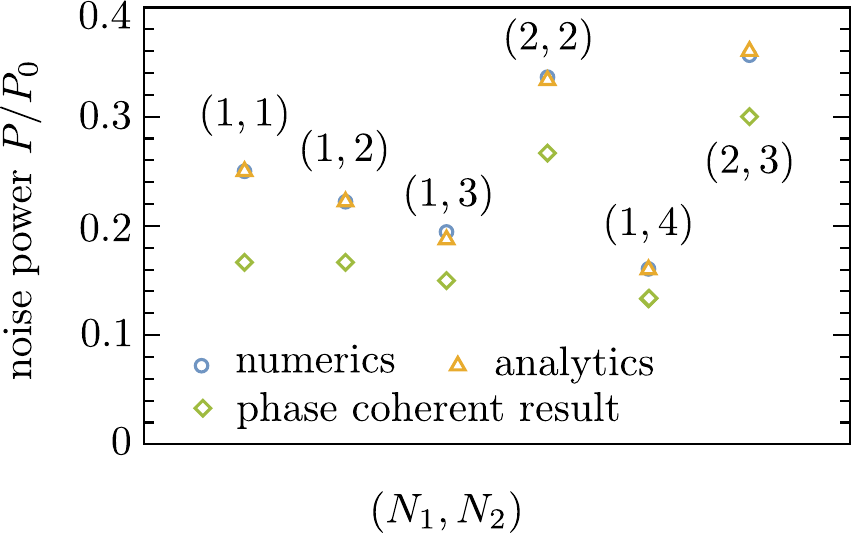}}
\caption{Noise power for the partitioning of $N$ modes into $N_1+N_2$ modes, for different values of $N_1,N_2$. The numerical results are computed from Eq.\ \eqref{Pnoiseformulasumps}, for a single realization of the unitaries (no averaging), at $\varepsilon=0.8$ and ${\cal L}=10$. The analytical data points are the result \eqref{PnoiseFano} for strong dephasing, which are a factor $1+1/N$ larger than the ensemble-averaged phase-coherent result \eqref{Pcoherent}.
}
\label{fig_noiseresults}
\end{figure}

We now make two observations. The first is that the eigenvalue statistics of the transfer probability matrix ${\cal T}_{\bm s}$ for a single quantum trajectory, in the ensemble of random unitary matrices, is independent of the measurement outcome or measurement strength. The proof is similar to the single-mode case considered above, see App.\ \ref{app_calTsdistribution}. This implies for the ensemble average $\mathbb{E}$ that
\begin{subequations}
\begin{align}
&\mathbb{E}[\operatorname{Tr}{\cal T}_{\bm s}]=\mathbb{E}[G_{\rm coh}/G_0],\\
&\mathbb{E}[\operatorname{Tr}{\cal T}_{\bm s}(1-{\cal T}_{\bm s})]=\mathbb{E}[ P_{\rm coh}/P_0],\\
&\mathbb{E}[(\operatorname{Tr}{\cal T}_{\bm s})^2]-\mathbb{E}[\operatorname{Tr}{\cal T}_{\bm s}]^2={\rm Var}\,(G_{\rm coh}/G_0),
\end{align}
\end{subequations}
where $G_{\rm coh}$ and $P_{\rm coh}$ are the phase coherent conductance and noise power ($\varepsilon=0$).

The second observation is that for strong dephasing, in the limit $\varepsilon^2{\cal L}\rightarrow\infty$, the quantum trajectories become independent and the sum over the $2^{\cal L}$ measurement outcomes self-averages to the expectation value $\mathbb{E}$ in the unitary ensemble. Combining the two observations, we conclude that
\begin{align}
P_{\rm noise}/P_0={}&\mathbb{E}[\operatorname{Tr}{\cal T}_{\bm s}(1-{\cal T}_{\bm s})]+\mathbb{E}[(\operatorname{Tr}{\cal T}_{\bm s})^2]-\mathbb{E}[\operatorname{Tr}{\cal T}_{\bm s}]^2\nonumber\\
={}&\mathbb{E}[ P_{\rm coh}/P_0]+{\rm Var}\,(G_{\rm coh}/G_0).\label{Pnoiseanalytics}
\end{align}
This is a remarkable relation between a transport property of a single system with strong dephasing, on the left-hand-side, and ensemble-averaged phase-coherent transport properties, on the right-hand-side.

For full phase coherence, one has the CUE relation \cite{Sav06} $\mathbb{E}[P_{\rm coh}/P_0]=N\operatorname{Var}(G_{\rm coh}/G_0)$. We thus obtain without further calculation the strong dephasing limits of noise power $P_{\rm noise}$ and Fano factor $F$,
\begin{equation}
\begin{split}
&P_{\rm noise}=\frac{N+1}{N}\mathbb{E}[P_{\rm coh}]=P_0\frac{N_1^2 N_2^2}{N^2(N-1)},\\
&F=\frac{P_{\rm noise}}{eVG}=\frac{N_1N_2}{N(N-1)},
\end{split}
\label{PnoiseFano}
\end{equation}
where we have substituted Eq.\ \eqref{Pcoherent} and $G/G_0=N_1N_2/N$. In Fig.\ \ref{fig_noiseresults} we check that this simple formula agrees well with a numerical simulation, without the need for ensemble averaging.

\textbf{\em Conclusion ---}  Experiments \cite{Mat15,Kum15} on the partition noise in a graphene \textit{p-n} junction have been analyzed in terms of a semiclassical formula \cite{Aba07} which is a factor $1-1/N^2$ \textit{smaller} than the phase coherent result \eqref{Pcoherent}. Our pure dephasing result, in the regime $L\gg L_\phi >L_{\rm mixing}$, is \textit{larger} by a factor $1+1/N$. For the experimental case $N_1=2=N_2$ (lowest Landau level with strong spin-orbit scattering), this implies a phase coherent Fano factor of 4/15, in between the semiclassical value 1/4 and the strong dephasing value of 1/3. The experimental values \cite{Mat15,Kum15} are well below these values, presumably because of inelastic scattering.

From a conceptual point of view, our finding that pure dephasing can increase the noise power is unexpected, but it has a simple intuitive interpretation in terms of self-averaging quantum trajectories. Earlier theoretical studies \cite{Che11,Ma18} of dephasing in the \textit{p-n} junction were based on the dephasing-probe model of decoherence \cite{For07}. This model suppresses both sample-to-sample fluctuations and time-dependent fluctuations (shot noise) in the current. While fully applicable to decoherence by inelastic scattering, pure dephasing only suppresses the sample-to-sample fluctuations, preserving and even enhancing the time-dependent fluctuations.

\textbf{\em Acknowledgments ---}
This work was supported by the Netherlands Organisation for Scientific Research (NWO/OCW), as part of Quantum Limits (project number {\sc summit}.1.1016).


\appendix


\section{Calculation of the transfer probability}
\label{app_secondmoment}

{\tt\small Comment: In an \href{https://arxiv.org/abs/2509.10242v1}{earlier version} of this appendix, I calculated the second moment of the transfer probability from recursion relations for the quantum trajectories. I needed to make the large-${\cal L}$ approximation that neglects correlations between the binary strings of measurement outcomes in different quantum trajectories. In the present version of the appendix I avoid that approximation, starting from a recursion relation for the full transfer probability, summed over all quantum trajectories. This new approach is both exact and more efficient, allowing the calculation of all moments of the transfer probability.}

\subsection{Recursion relation}

The rescaled Kraus matrix \eqref{KSrelation} for $N=2$ is given by the matrix product
\begin{equation}
{\cal S}_{\bm s}=U_{{\cal L}}e^{s_{\cal L}\sigma_z\gamma/2}{U}_{{\cal L}-1}e^{s_{{\cal L}-1}\sigma_z\gamma/2}{U}_{{\cal L}-2}{\cdots {U}_1 e^{s_{1}\sigma_z\gamma/2}}{U}_0.\label{KrausmatrixcalS}
\end{equation}
The weak measurement is assumed to be of mode number 1, and we have substituted $e^{-s_\ell\gamma/2}e^{s_{\ell}\gamma|1\rangle\langle 1|}=e^{s_{\ell}\sigma_z\gamma/2}$. The binary string $\bm{s}=\{s_{\cal L},s_{{\cal L}-1},\ldots s_1\}$, $s_n=\pm 1$, labels a quantum trajectory. The matrices $U_n$ are a set of independent uniformly distributed $2\times 2$ unitary matrices, the same set for each of the $2^{\cal L}$ quantum trajectories.

The transfer probability ${\cal T}$ follows from
\begin{equation}
{\cal T}=C_{\cal L}\textstyle{\sum_{\bm s}} |t_{\bm s}|^2,\;\;C_{\cal L}=(2\cosh\gamma)^{-\cal L},\;\;t_{\bm s}=({\cal S}_{\bm s})_{21}.
\end{equation}
We may equivalently write this as
\begin{equation}
{\cal T}=\operatorname{Tr}P_-{\cal G},\;\;{\cal G}=C_{\cal L}\sum_{\bm s}{\cal S}_{\bm s}^{\vphantom{\dagger}}P_+{\cal S}_{\bm s}^\dagger,\;\;P_\pm=\tfrac{1}{2}(1\pm \sigma_z). 
\end{equation}

Since
\begin{equation}
e^{s\sigma_z\gamma/2}=\cosh(\gamma/2)+s\sigma_z\sinh(\gamma/2),
\end{equation}
we have upon summing over $s_{{\cal L}+1}=\pm 1$ the iteration
\begin{align}
{\cal G}({\cal L}+1)={}&\frac{\cosh^2(\gamma/2)}{\cosh\gamma}U_{{\cal L}+1}^{\vphantom{\dagger}}{\cal G}({\cal L})U_{{\cal L}+1}^\dagger\nonumber\\
&+\frac{\sinh^2(\gamma/2)}{\cosh\gamma}U_{{\cal L}+1}^{\vphantom{\dagger}}\sigma_z{\cal G}({\cal L})\sigma_zU_{{\cal L}+1}^\dagger.
\end{align}
The iteration starts with
\begin{equation}
{\cal G}(0)=U_0P_+U_0^\dagger.
\end{equation}
The iteration is trace preserving,
\begin{equation}
\operatorname{Tr}{\cal G}({\cal L}+1)=\operatorname{Tr}{\cal G}({\cal L})\Rightarrow \operatorname{Tr}{\cal G}({\cal L})=1.
\end{equation}

\subsection{Bloch vector representation}

The $2\times 2$ trace-one Hermitian matrix ${\cal G}$ has the parametrization
\begin{equation}
{\cal G}=\tfrac{1}{2} \sigma_0 +\tfrac{1}{2}\bm{g}\cdot\bm{\sigma},\;\;\bm{g}\in\mathbb{R}^3,
\end{equation}
corresponding to the transfer probability
\begin{equation}
{\cal T}=\tfrac{1}{2}+\tfrac{1}{2}g_z.
\end{equation}

The recursion relation for the Bloch vector $\bm{g}$ is
\begin{equation}
\bm{g}({{\cal L}+1})=\frac{1}{\cosh\gamma}R(U_n)\Omega \bm{g}({\cal L}),\;\;\Omega=\begin{pmatrix}
1&0&0\\
0&1&0\\
0&0&\cosh\gamma
\end{pmatrix},
\end{equation}
where $R(U)$ is the three-dimensional rotation matrix corresponding to the unitary matrix $U$,
\begin{equation}
R_{ij}(U)=\tfrac{1}{2}\operatorname{Tr}(\sigma_i U\sigma_j U^\dagger),\;\;
U\sigma_j U^\dagger=\sum_{i=1}^3 R_{ij}(U)\sigma_i.
\end{equation}

At each step of the iteration, the direction $\hat{n}$ of the Bloch vector $\bm{g}=r\hat{n}$ is uniformly distributed on the unit sphere. The length $r=|\bm{g}|$ has the recursion relation
\begin{equation}
r({{\cal L}+1})^2=\frac{1}{\cosh^2\gamma}r({\cal L})^2[1+n_z^2\sinh^2\gamma ],
\end{equation}
with $n_z$ uniformly distributed in $(-1,1)$, independently of $r$. Starting from $r(0)=1$, we have
\begin{equation}
r({\cal L})^2=(\cosh\gamma)^{-2{\cal L}}\prod_{i=1}^{\cal L}[1+u_i^2\sinh^2\gamma ],\label{r2result}
\end{equation}
with $u_1,u_2,\ldots u_{\cal L}$ a set of independent random variables uniformly distributed in $(-1,1)$. The transfer probability is
\begin{equation}
{\cal T}({\cal L})=\tfrac{1}{2}+\tfrac{1}{2}r({\cal L})u_0,\label{Tu0relation}
\end{equation}
with one more independent uniform variable $u_0\in(-1,1)$.

\subsection{Moments}

Upon averaging over the $u_n$'s we find from Eqs.\ \eqref{r2result} and \eqref{Tu0relation} that $\mathbb{E}[({\cal T}-1/2)^p]=0$ for $p$ odd, while for $p$ even
\begin{align}
\mathbb{E}[({\cal T}-1/2)^p]={}&2^{-p}(1+p)^{-1}(\cosh\gamma)^{-p{\cal L}}\nonumber\\
&\times\left[ \sum_{n=0}^{p/2}{{p/2}\choose n}\frac{\sinh^{2n}\gamma}{2n+1}\right]^{\cal L}.
\end{align}

In particular, the mean $\mathbb{E}[{\cal T}]=1/2$, independent of ${\cal L}$, and the variance is
\begin{align}
\operatorname{Var}[{\cal T}]={}&\tfrac{1}{12}(\cosh\gamma)^{-2{\cal L}}\left(1+\tfrac{1}{3}\sinh^2\gamma\right)^{\cal L}\nonumber\\
={}&\tfrac{1}{12}\left(1-\tfrac{4}{3}\varepsilon^2+\tfrac{2}{3}\varepsilon^4\right)^{\cal L},
\end{align}
tending to $\tfrac{1}{12}e^{-4\varepsilon^2{\cal L}/3}$ in the limit $\varepsilon\rightarrow 0$, ${\cal L}\rightarrow\infty$ at fixed $\varepsilon^{2}{\cal L}$.

\subsection{Full distribution}
\label{appA_full}

For $\varepsilon\ll 1$ one has from Eq.\ \eqref{r2result}
\begin{equation}
\begin{split}
&\ln r=\varepsilon^2 \sum_{i=1}^{\cal L}(u_i^2-1)+{\cal O}(\varepsilon^4),\\
&\Rightarrow \mathbb{E}[\ln r]=-\tfrac{2}{3}\varepsilon^2{\cal L},\;\;{\rm Var}[\ln r]=\tfrac{4}{45}\varepsilon^4{\cal L},
\end{split}
\end{equation}
producing for $\varepsilon\rightarrow 0$, ${\cal L}\rightarrow\infty$, at fixed $\varepsilon^2{\cal L}$, a delta-function distribution of the length $r$ of the Bloch vector, centered at $r_{\rm peak}=e^{-2\varepsilon^2{\cal L}/3}$. The width of the distribution of $r$  vanishes in this limit.

\begin{figure}[tb]
\centerline{\includegraphics[width=0.8\linewidth]{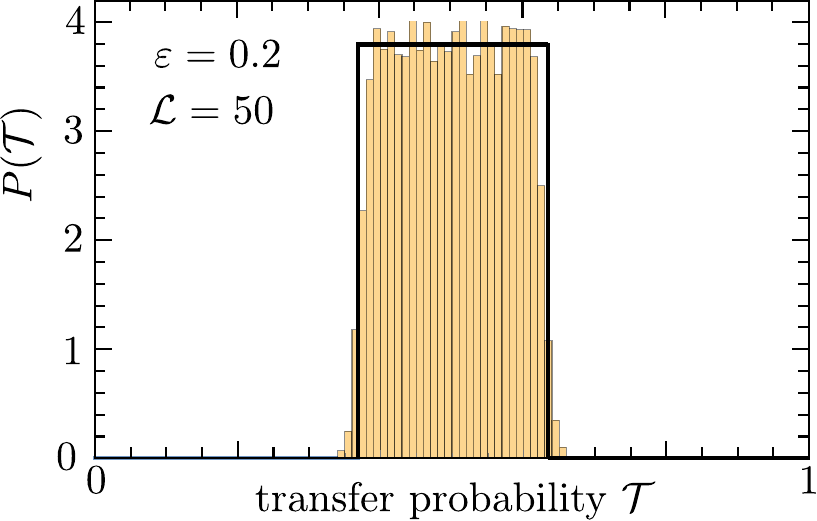}}
\caption{Distribution of the transfer probability in the ensemble of random unitary matrices [Haar-measure distributed in ${\rm U}(N)$ with $N=2$]. The histogram is computed numerically, averaged over $10^4$ sets of unitaries $U_0,U_1,\ldots U_{\cal L}$, for $\varepsilon=0.2$ and ${\cal L}=50$. The black line is the uniform distribution \eqref{PcalTuniform} in the small-$\varepsilon$, large-${\cal L}$ limit. This figure can be compared with Fig.\ \ref{fig_ptransfer} from the main text, for shorter ${\cal L}$ and larger $\varepsilon$, when the distribution is more rounded. 
}
\label{fig_uniformdistribution}
\end{figure}

Eq.\ \eqref{Tu0relation} then implies that the small-$\varepsilon$, large-${\cal L}$ limit of the distribution of $(2{\cal T}-1)/r_{\rm peak}$ is uniform in the interval $(-1,1)$, hence
\begin{equation}
P({\cal T})\rightarrow e^{2\varepsilon^2{\cal L}/3}\Theta\bigl(e^{-2\varepsilon^2{\cal L}/3}-|1-2{\cal T}|\bigr),\label{PcalTuniform}
\end{equation}
with $\Theta(x)$ the unit step function. In Fig.\ \ref{fig_uniformdistribution} we compare this analytical result with the numerics.

\section{Transfer statistics of a single quantum trajectory}
\label{app_calTsdistribution}

In the main text we showed, for the case $N=2$, $N_1=N_2=1$, that the statistics of the transfer probability ${\cal T}_{\bm s}$ for a single quantum trajectory is independent of the measurement outcome or measurement strength. The proof for arbitrary $N$ and arbitrary $N_1,N_2$ is similar. The object ${\cal T}_{\bm s}$, defined in Eq.\ \eqref{FxipsTs}, is then an $N_2\times N_2$ matrix, and we will be considering the statistics of its eigenvalues.

To find the eigenvalue statistics of ${\cal T}_{\bm s}$, we again use that ${\cal S}_{\bm s}$ can be factored as ${\cal S}_{\bm s}=UM$ with $U$ uniformly distributed in ${\rm U}(N)$, independently of the $N\times N$ complex matrix $M$. Place the first $N_1$ columns of $M$ in an $N\times N_1$ matrix $\delta M$. We assume that $\delta M$ is not rank-deficient, that it has rank $N_1$. In this generic case $\delta M=QX$ has the QR decomposition with an invertible $N_1\times N_1$ matrix $X$ and an $N\times N_1$ matrix $Q$ with orthonormal columns, $Q^\dagger Q=I$.

The $N\times N_1$ matrix $V=UQ$ also has orthonormal columns. Partition $V$ into two halves,
\begin{equation}
V=\begin{pmatrix}
V_{\rm top}\\
V_{\rm bottom}
\end{pmatrix},\;\;V^\dagger V=I,
\end{equation}
with $V_{\rm top}$ of dimension $N_1\times N_1$ and $V_{\rm bottom}$ of dimension $N_2\times N_1$. The reflection and transmission blocks of ${\cal S}_{\bm s}$, partitioned as in Eq.\ \eqref{Spartioning}, are then given by $r_{\bm s}=V_{\rm top}X$, $t_{\bm s}=V_{\rm bottom}X$.

We now use that
\begin{align}
r_{\bm s}^\dagger r_{\bm s}^{\vphantom{\dagger}}+t_{\bm s}^\dagger t_{\bm s}^{\vphantom{\dagger}}={}&X^\dagger V_{\rm top}^\dagger V_{\rm top}^{\vphantom{\dagger}}X+X^\dagger V_{\rm bottom}^\dagger V_{\rm bottom}^{\vphantom{\dagger}}X\nonumber\\
={}&X^\dagger V^\dagger VX=X^\dagger X,
\end{align}
hence
\begin{align}
{\cal T}_{\bm s}={}&(r_{\bm s}^\dagger r_{\bm s}^{\vphantom{\dagger}}+t_{\bm s}^\dagger t_{\bm s}^{\vphantom{\dagger}})^{-1}t_{\bm s}^\dagger t_{\bm s}^{\vphantom{\dagger}}\nonumber\\
={}&(X^\dagger X)^{-1}X^\dagger V_{\rm bottom}^\dagger V_{\rm bottom}^{\vphantom{\dagger}}X\nonumber\\
={}&X^{-1}V_{\rm bottom}^\dagger V_{\rm bottom}^{\vphantom{\dagger}}X.
\end{align}

The matrix ${\cal T}_{\bm s}$ is therefore related to the Hermitian matrix $V_{\rm bottom}^\dagger V_{\rm bottom}^{\vphantom{\dagger}}$ by a similarity transformation, so these two matrices have the same eigenvalues. Moreover, since the Haar measure of $U$ is invariant under multiplication by an arbitrary fixed unitary, we may choose a basis such that $Q$ contains the first $N_1$ columns of the unit matrix, and then $V_{\rm bottom}$ is just the lower-left block of $U$. All dependence on the measurements has dropped out.

\section{Log-normal distribution of the Born weights}
\label{app_lognormal}

\begin{figure}[tb]
\centerline{\includegraphics[width=0.9\linewidth]{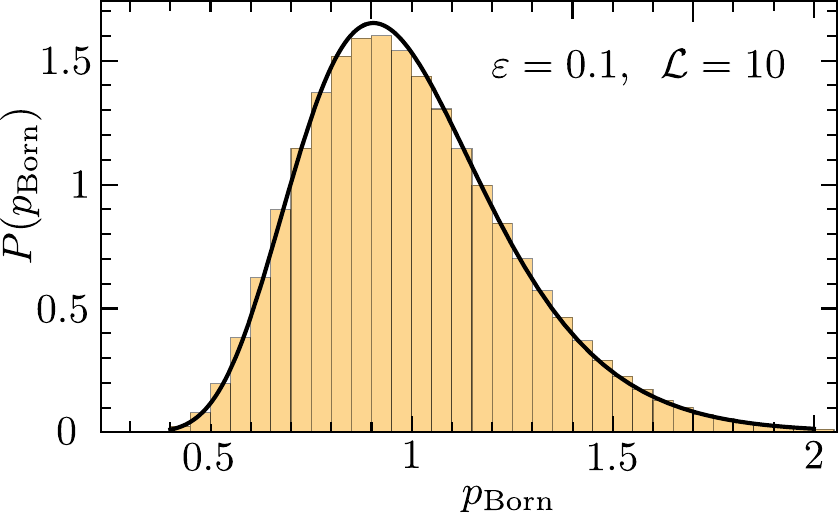}}
\caption{
Probability distribution function of the rescaled Born weight $p_{\rm Born}=2^{\cal L}p_{\bm s}$, for ${\cal L}=10$ measurements of strength $\varepsilon=0.1$. The histogram is computed numerically from the Kraus matrix \eqref{Krausmatrixdef} (averaged over $10^4$ sets of unitaries), the solid curve is the log-normal distribution \eqref{PpBorn}, obtained in the weak measurement limit.
}
\label{fig_check_appB}
\end{figure}

The results from this appendix are not used in the main text, it is provided as additional information.

We compute the probability distribution function of the Born weights $p_{\bm s}$ for $N=2$ modes, in the weak measurement limit $\varepsilon\rightarrow 0$, ${\cal L}\rightarrow\infty$ at fixed $\varepsilon^2{\cal L}$. For that purpose we work with the rescaled weight
\begin{equation}
p_{\rm Born}=2^{\cal L}p_{\bm s}=(\cosh\gamma)^{-{\cal L}}\left\|{\cal S}_{\bm s}\cdot\textstyle{1\choose 0}\right\|^2,
\end{equation}
which averages to unity, independent of ${\cal L}$. For later use we note that
\begin{equation}
(\cosh\gamma)^{-{\cal L}}=(1-\varepsilon^2)^{{\cal L}}\rightarrow e^{-\varepsilon^2{\cal L}}
\end{equation}
in the weak measurement limit.

Consider the action of the Kraus matrix \eqref{KrausmatrixcalS} on the spinor ${1\choose 0}$ and denote by $\Psi_n=\|\Psi_n\|{\psi_1\choose \psi_2}$ the spinor just before the $n$-th factor $e^{s_n(\gamma/2)\sigma_z}$. The spinor ${\psi_1\choose \psi_2}$ has unit norm and is isotropically distributed. The log-norm increment is
\begin{align}
\xi_n={}&\ln\| e^{s_n(\gamma/2)\sigma_z}\Psi_n\|-\ln\|\Psi_n\|\nonumber\\
={}&\tfrac{1}{2}\ln(e^{s_n\gamma}|\psi_1|^2+e^{-s_n\gamma}|\psi_2|^2)\nonumber\\
={}& \tfrac{1}{\sqrt 2}s_n\varepsilon\Delta + \tfrac{1}{2}\varepsilon^2(1-\Delta^2) + {\cal O}(\varepsilon^3),
\end{align}
with
\begin{equation}
\Delta = |\psi_1|^2-|\psi_2|^2=2|\psi_1|^2-1 \in (-1,1).
\end{equation}
We have used that $\gamma=\sqrt{2}\,\varepsilon+{\cal O}(\varepsilon^3)$.

The distribution of $|\psi_1|^2$ is uniform on $(0,1)$ \cite{note4}, hence
\begin{equation}
\mathbb{E}[\Delta] = 0, \;\; \mathbb{E}[\Delta^2] = \tfrac13.
\end{equation}
Averaging over $\Delta$ and $s_n=\pm 1$ gives
\begin{equation}
\mathbb{E}[\xi] = \tfrac{1}{3}\,\varepsilon^2 + {\cal O}(\varepsilon^3),\;\;
\mathrm{Var}[\xi] =  \tfrac{1}{6}\varepsilon^2 + {\cal O}(\varepsilon^3).
\end{equation}

Subsequent increments $\xi_n$ are independent. By the central limit theorem, the probability distribution of $ \sum_{n=1}^{\cal L} \xi_n $ thus tends to a Gaussian for large ${\cal L}$, with mean $\tfrac{1}{3}\varepsilon^2{\cal L}\equiv\tau$ and variance $\tau/2$. 

The Born weight 
\begin{equation}
p_{\rm Born}=e^{-\varepsilon^2{\cal L}}\exp\left(2\sum_{n=1}^{\cal L} \xi_n\right)
\end{equation}
therefore has a log-normal distribution with $\mathbb{E}[\ln p_{\rm Born}]=-\tau$ and $\operatorname{Var}[\ln p_{\rm Born}]=2\tau$,
\begin{equation}
P(p_{\rm Born})=\frac{1}{p_{\rm Born}\sqrt{4\pi\tau}}\exp\left[-\frac{(\ln p_{\rm Born}+\tau)^2)}{4\tau}\right].\label{PpBorn}
\end{equation}
One finds that $\mathbb{E}[p_{\rm Born}]=1$, $\operatorname{Var}p_{\rm Born}=e^{2\tau}-1$. In Fig.\ \ref{fig_check_appB} we check agreement with the numerics.


\begin{thebibliography}{99}
\bibitem{notedephasing} As emphasised by Marquardt and Bruder \cite{Mar04}, one should distinguish pure dephasing from ``phase averaging'', either in the form of thermal averaging (when $k_{\rm B}T\gg eV$), or due to a slowly varying environment (when $\hbar\omega_c\ll eV$). A fully quantum mechanical treatment is only needed for pure dephasing.
\bibitem{note5} In Eq.\ \eqref{Fxidetrt} we used the identities $e^{\xi P_{\rm out}}=1+(e^\xi-1)P_{\rm out}$, $\delta^{2{\cal L}} e^{\gamma {\cal L}}=(2\cosh\gamma)^{-\cal L}$ and $\operatorname{Det}[1+P_{\rm in}(cX-1)]=c^{N_1}\operatorname{Det}[1+P_{\rm in}(X-1)]$, valid for any scalar $c$ and matrix $X$.
\bibitem{note4} Uniformity on $(0,1)$ of $|U_{ij}|^2$ when $U$ is uniformly distributed in ${\rm U}(2)$ can be seen, for example, by taking $i=j=1$ and noting that a column vector $(U_{11},U_{21})^\top$ of $U$ corresponds to a point on the Bloch sphere, with polar angle $\theta$ and $|U_{11}|^2=\cos^2(\theta/2)$. The measure $\sin\theta\, d\theta$ on the Bloch sphere then gives the uniform distribution of $|U_{11}|^2$, and by symmetry also for other choices of indices.
\bibitem{note_numerics} The numerical histogram for ${\cal T}$ in Fig.\ \ref{fig_ptransfer} has mean 0.499 and variance 0.00266, close to the analytical results of, respectively, 0.5 and 0.00265 [see Eq.\ \eqref{VarTresult}]. This is for a relatively large $\varepsilon$, for weaker dephasing the distribution of ${\cal T}$ has a block form, uniform in a narrow interval around $1/2$, see App.\ \ref{appA_full}.
\end{thebibliography}
\end{document}